\input lecproc.cmm

\contribution{Kelvin-Helmholtz instability of relativistic jets
-- the transition from linear to nonlinear regime}
\contributionrunning{Kelvin-Helmholtz instability of relativistic jets}
\author{Micha{\l} Hanasz}
\address{Centre for Astronomy,
Nicolaus Copernicus University,  87-148 Piwnice/Toru{\'n}}

\abstract{
The observed wiggles and knots in astrophysical jets as well as the curvilinear
motion of radio emitting features are frequently interpreted as signatures of
the Kelvin--Helmholtz (KH) instability (eg. Hardee 1987).
We investigate the KH instability of a hydrodynamic jet composed of
a relativistic gas, surrounded by a nonrelativistic external medium and moving
with a relativistic bulk speed.
We show basic nonlinear effects, which become important for a finite
amplitude KH modes. Since the KH instability in supersonic jets involves
acoustic waves over-reflected on jet boundaries, the basic nonlinear effect
relies on the steepening of the acoustic wave fronts, leading to the formation
of shocks.  It turns our that the shocks appear predominantly in
the external nonrelativistic gas, while the internal acoustic waves remain
linear for a much longer time. In addition, the external medium "hardens"
as soon as the boundary oscillation velocity becomes comparable to the external sound
speed. On the other hand, the amplification of internal
waves due to the over-reflection is limited by a nonlinearity of the
Lorentz $\gamma$ factor.  This implies that the sidereal oscillations of the jet
boundary, resulting from the K-H instability, are limited to very small
amplitudes comparable to a fraction of the
jet radius.
}

\titlea{1}{Introduction}

The Kelvin-Helmholtz (K-H) instability of supersonic jets
excites oblique acoustic waves in jets and the ambient medium.  Internal
acoustic waves are amplified by multiple reflection from jet boundaries with the
absolute value of the reflection coefficient ${\cal R} = (Z_e-Z_i)/(Z_e+Z_i) $
larger than 1 (Payne and Cohn 1985), where $Z_e= P_e/V_e$, $Z_i=P_i/V_i$ are
the complex acoustic
impedances and $P_i$, $P_e$, $V_i$, $V_e$ are the pressure and velocity
oscillation amplitudes for acoustic waves in the internal and external gases.
This effect is called over-reflection and is caused by the Bernouli effect.
Acoustic impedance tells us if the medium is 'soft' or 'hard' depending on what
pressure oscillation amplitude is necessary to force the unit velocity
oscillation amplitude.  The contact surface of internal
 and external gases
undergoes a wavelike deformation, emitting outward propagating acoustic waves
in the external medium.

We apply the relativistic equations
of hydrodynamics and the relativistic equation of state, with the adiabatic
index $\Gamma_i=4/3$ and internal sound speed $c_{si} = c/\sqrt{3}$,
for the jet gas. The external gas is described by the
nonrelativistic equations of hydrodynamics
and the nonrelativistic equation of state with $\Gamma_e = 5/3$.
The boundary conditions ensure that pressure and transversal
displacements of the internal and external gases are equal at the jet boundary.
In the present considerations we assume 2D slab geometry.

The linearized theory predicts a variety of modes of the K-H instability,
classified with respect to their symmetry, longitudinal wavenumber, azimuthal
wavenumber and the number of nodes within the jet radius.  There is a
qualitative correspondence between symmetric (antisymmetric)
modes of 2D jets and pinching (helical) modes of cylindrical
jets.

The main problem of the linearized theory of the K-H instability is such that it is
valid only for infinitesimally small amplitude perturbations.  In the next
sections we shall investigate some aspects of the nonlinear development of the
K-H instability in relativistic, supersonic jets.

\titlea{2}{Nonlinear steepening of acoustic waves}

Since the acoustic wave amplitude grows due to the multiple over-reflection, one
can expect that sooner or later the nonlinear range will be reached.  Observing
waves in their direction of propagation we can tentatively apply a one
dimensional description.  The nonlinear propagation of sound waves in one
dimension is analytically described within the theory of simple waves, for
which the Riemann invariants allow to specify explicitly nonlinear analytical
relations between pressure, density, sound speed and gas velocity (see eg.
Landau and Lifshitz (1959) for the classical case and Anile (1989) for the
relativistic case).  The essential property of nonlinear acoustic waves is such
that the propagation speed is dependent on density, what leads to their
steepening and formation of shocks.

It is relatively easy to calculate time and the path lengths which is necessary
to form shock fronts from an initially sinusoidal wave of a given amplitude.
It turns out that for a given mode of K-H instability and perturbation
amplitude, the path length measured in the direction perpendicular to the
jet axis, necessary for the formation of shocks  is an order of
magnitude larger in the relativistic jet gas than in the external medium.

Assuming $\gamma=10$ for the present considerations, the corresponding time
necessary to form shocks (measured in the reference frame of external gas)
is two orders of magnitude larger in the internal gas
due to the additional $\gamma$ factor .
Thus, we expect that for given amplitude of jet boundary oscillations due to
K-H instability, the relativistic jet gas
should be much more ``smooth'' than the nonrelativistic external medium or
the nonrelativistic jet gas.

\titlea{3}{Nonlinear reaction of external medium}

The formalism of simple acoustic waves applied to the nonrelativistic external
gas with the adiabatic index $\Gamma_e = 5/3$ leads to the following relation
between pressure and velocity (for details see Landau and Lifshitz 1959)

$$
p_e = p_{e0} \left( 1 + {{1}\over{3}} {{v_e}\over{c_{se0}}}\right)^5
$$
It follows from the above relation that the internal gas feels more and more
hard interface, while the perturbation amplitude grows.  This means that the
external acoustic impedance grows, i.e. one needs much more effort to increase
the velocity oscillations of the boundary when the perturbation amplitude
is large.
The effect becomes important
when the jet boundary starts to oscillate with the speed comparable to the
external sound speed.  A more detailed considerations show, however, that this
effect does not limit definitely the displacement amplitude of the contact
surface.

\titlea{4}{Reflection of internal acoustic waves}

Let us focus our attention on reflection of internal acoustic waves, observed
in the reference frame comoving with the unperturbed flow. Since
the formation of shocks is very slow in the
relativistic internal gas, we shall assume that the fundamental frequency of
acoustic waves dominates and the amplitudes of higher harmonics are small.

In the linear range the typical reflection coefficients for the most unstable
modes are of the order of 3-10 (eg. for
$\gamma=10$, $\nu=\rho_i/\rho_e=0.01$).  It is obvious, however, that we can
not magnify the amplitude of sound waves to an unlimited value during the
multiple reflection because of the relativistic limitation of velocity.

Analyzing internal sound waves in the vicinity of a given point on the jet's
boundary,
we have to take into considerations the sum of velocity perturbations
corresponding to the incident and reflected waves. The relativistic velocity
limitation can be roughly written for the parallel component of velocity
perturbation

$$
\left|v_{\parallel}^{(j)+} + v_{\parallel}^{(j)-} \right| < c,
$$
where both the incident $v_{\parallel}^{(j)+}$ and reflected $v_{\parallel}^{(j)-} $
components have the same directions and the superscript $(j)$ means the reference
frame comoving
with the jet velocity. If the parallel velocity oscillations corresponding
to the incident wave have already
a large amplitude
$V_{\parallel}^{(j)+} =c/2$, then the amplitude $V_{\parallel}^{(j)-} $ of the
reflected wave has to be smaller than $c/2$. Then we have

$$
\mid {\cal R} \mid = {{\mid V_{\parallel}^{(j)-} \mid}\over{\mid
V_{\parallel}^{(j)+} \mid}} < 1
$$
This means that there is no more amplification of acoustic waves above a
certain amplitude  of the incident
wave (which is in fact smaller than $c/2$). {\it
Therefore we expect the relativistic saturation of the K-H instability.} This
effect is caused by the nonlinearity contained in the Lorentz factor $\gamma$
in the relativistic Euler equation.  The nonlinearity prevents the gas velocity
to exceed the speed of light.

Let us estimate the boundary oscillation amplitude corresponding to
$V_{\parallel}^{(j)\pm} \sim c/2$. The Lorentz transformation implies

$$
V_{\parallel}^{(j)} = \gamma^2 V_{\parallel}^{(e)},
$$
$$
V_{\perp}^{(j)} = \gamma V_{\perp}^{(e)},
$$
where $V_{\parallel}^{(j)}$, $V_{\parallel}^{(e)}$
$V_{\perp}^{(j)}$ and $V_{\perp}^{(e)}$ are perturbations of velocity parallel
and perpendicular to the jet axis, measured in the reference frames of jet $(j)$
and the external gas $(e)$.

The linear solution allows to figure out that $\mid V_{\parallel}^{(j)\pm}\mid\sim
\mid V_{\perp}^{(j)\pm}\mid$,
then the limitation $\mid V_{\parallel}^{(j)\pm} \mid$ to $c/2$ in the jet
reference frame implies the corresponding limitation in the reference frame
of external gas

$$
V_{\parallel}^{(e)\pm} < {{c}\over{2\gamma^2}}
$$
$$
V_{\perp}^{(e)\pm} < {{c}\over{2\gamma}}
$$
Let us take into account a typical linear solution
$k\sim 1/R$, $\omega_r \sim c_{si}/R$, $\omega_i \sim 0.1 c_{si}/R$.
These parameters are in fact dependent on the mode of perturbation, but we
shall neglect that for simplicity.
One can deduce that the maximal displacement amplitude $A$ of the jet boundary
is

$$
A \sim {{V_{\perp}^{(e)\pm}}\over{|\omega|}} \sim  {{R}\over{\gamma}},
$$
where $R$ is the jet radius and $\omega$ is the complex frequency.
It is apparent that even if the internal gas oscillates with velocity
approaching the speed of light (in the jet reference frame), the external
boundary
oscillates with a very small amplitude of the order of $R/\gamma$.

\titlea{5}{Conclusions}

\smallskip
\item{1.} The K-H instability saturates for very small amplitudes of lateral
displacements of jet's boundaries. This saturation is caused by the
relativistic limitation of velocity of the jet gas
and is additionally supported by
the nonlinear reaction of the external medium.
\item{2.} The patterns of internal oblique shocks form very slowly in the
internal relativistic gas.
\item{3.} The results obtained by means of the
multiple timescale method, applied to the K-H instability in
relativistic jets, lead to the same conclusions (Hanasz, 1995).
\item{4.} Our results are consistent  with the results of
numerical simulations of relativistic jets (Marti et al. 1997,
Duncan and Hughes 1994), where the lack of internal jet structure is observed
in the case of relativistic equation of state of the jet gas.
\item{5.} It seems to be unlikely that the curvilinear motion of superluminal
knots in some jets (like 3C 345 or 3C 273) is caused by
the Kelvin--Helmholtz instability.

\begrefchapter{References}
\ref Anile, A.M., 1989, {\it Relativistic Fluids and Magneto--Fluids},
Cambridge University Press
\ref Duncan, G.C., Hughes, P.A., 1994 ApJ 436, L119
\ref Hanasz, M., 1995, PhD Thesis, Nicolaus Copernicus University,
Toru\'n
\ref Hardee, P.E., 1987, ApJ 334, 70
\ref Landau, L.D., Lifshitz, E.M., 1959, {\it Fluid Mechanics}, Pergamon
Press
\ref Marti, M.A., M\"uller, E., Font, J.A., Ibanez, J.M., Marquina, A.,
1997, ApJ 479, 151
\ref Payne, D.G., Cohn,H., 1985, ApJ 191, 655
\endref

\byebye

\end{document}